\documentclass{article}
\usepackage{epsfig}

\newcommand\prl[3]   {{\it Phys.\ Rev.\ Lett.\ }{\bf #1} (#2) #3}
\newcommand\prd[3]   {{\it Phys.\ Rev.\ }{\bf D #1} (#2) #3}
\newcommand\zpc[3]   {{\it Z.\ Physik }{\bf C #1} (#2) #3}
\newcommand\npb[3]   {{\it Nucl.\ Phys.\ }{\bf B #1} (#2) #3}
\newcommand\plb[3]   {{\it Phys.\ Lett.\ }{\bf B #1} (#2) #3}
\newcommand\jhep[3]  {{\it J. High Energy Phys.\ }{\bf #1} (#2) #3}
\def\Frac#1#2{\frac{\displaystyle{#1}}{\displaystyle{#2}}}
\def\lsim{\raise0.3ex\hbox{$\;<$\kern-0.75em\raise-1.1ex\hbox{$\sim\;$}}}
\def\gsim{\raise0.3ex\hbox{$\;>$\kern-0.75em\raise-1.1ex\hbox{$\sim\;$}}}
\begin{document} 
\hyphenation{re-pha-sings bo-unds e-lec-tron u-ni-ver-sa-li-ty
sfer-mions in-de-pen-dent Yu-ka-wa se-arch char-gi-no
spec-ci-fically pha-ses model-in-de-pen-dent diago-na-lity 
                   an-ni-hi-la-tion pen-guin}
\thispagestyle{empty}
\begin{flushright}
SISSA/153/99/EP
\end{flushright}
\vskip 1cm
\begin{center}
{\Large \bf      %INSERT TITLE
Flavor Structure and Supersymmetric CP-Violation
\par} \vskip 2.em
{\large         %INSERT NAMES
{\sc  A. Masiero and O. Vives}  \\[1ex] %INSERT ADDRESS
{\em SISSA -- ISAS, via Beirut 2-4, 34013, Trieste, Italy and} \\
{\em INFN, Sezione di Trieste, Trieste, Italy}\\[1ex]
\par} 
\end{center} \par
\vskip 1cm 
\begin{abstract}
In this talk, we address the possibility of finding supersymmetry
through indirect searches in the $K$ and $B$ systems. We prove that, 
in the absence of the Cabibbo--Kobayashi--Maskawa phase, a 
general Minimal Supersymmetric Standard Model with all possible phases in 
the soft--breaking terms, but no new flavor structure beyond the usual 
Yukawa matrices, can never give a sizeable contribution to $\varepsilon_K$, 
$\varepsilon^\prime/\varepsilon$ or hadronic $B^0$ CP asymmetries. 
However, Minimal Supersymmetric models with additional flavor structures in 
the soft--supersymmetry breaking terms can produce large deviations from the 
Standard Model predictions. 
Hence, observation of supersymmetric contributions to CP asymmetries in B 
decays would be the first sign of the existence of new flavor structures in 
the soft--terms and would hint at a non--flavor blind mechanism of 
supersymmetry breaking.
\end{abstract}
\section{Introduction}
\label{sec:intro}

Beginning with its experimental discovery in $K$--meson decays, about 
three decades ago, the origin of CP violation has been one of the most 
intriguing questions in particle phenomenology. Notably, the subsequent 
experiments in the search for electric dipole moments (EDM) of the neutron and 
electron have observed no sign of new CP--violating effects despite their 
considerably high precision. Hence, the neutral $K$--system remains, so far, 
the only experimental information on the presence of CP--violation in nature.

In the near future, this situation will change.  
Not only the new $B$ factories will start measuring CP violation effects in 
$B^0$ CP asymmetries, but also the experimental sensitivity to the electric 
dipole moment of the neutron and the electron will be substantially 
improved. These new experiments will enlarge our knowledge of CP violation
phenomena and, hopefully, will show the existence of new sources of CP 
violation from models beyond the Standard Model (SM).

The Standard Model of electroweak interactions is known to be able to 
accommodate the experimentally observed CP--violation through a unique phase, 
$\delta_{CKM}$, in the Cabibbo--Kobayashi--Maskawa mixing matrix (CKM). 
However, most of the extensions of the SM include new observable phases
that may significantly modify the pattern of CP violation.
Supersymmetry is, without a doubt, one of the most popular extensions of the SM. 
Indeed, in the minimal supersymmetric extension of the SM (MSSM), there 
are additional phases which can cause deviations from the predictions of the 
SM. After all possible rephasings of the parameters and fields, there remain 
at least two new physical phases in the  MSSM Lagrangian. These phases can be 
chosen to be the phases of the Higgsino Dirac mass parameter 
($\varphi_{\mu}=\mbox{Arg}[\mu]$) and  the trilinear sfermion coupling 
to the Higgs, ($\varphi_{A_{0}}=\mbox{Arg}[A_{0}]$) 
\cite{2phases}. In fact, in the so--called Constrained Minimal Supersymmetric 
Standard Model (CMSSM), with strict universality at the Grand Unification 
scale, these are the only new phases present.

It was soon realized, that for most of the MSSM parameter space, the 
experimental bounds on the electric dipole moments of the electron and 
neutron constrained $\varphi_{A_0,\mu}$ to be at most ${\cal{O}}(10^{-2})$.
Consequently these new supersymmetric phases have been taken to vanish 
exactly in most studies in the framework of the MSSM.
  
However, in the last few years, the possibility of having non--zero SUSY phases
has again attracted a great deal of attention. Several new mechanisms have 
been proposed to suppress supersymmetric contributions to EDMs below the 
experimental bounds while allowing SUSY phases ${\cal{O}}(1)$. 
Methods of suppressing the EDMs 
consist of cancellation of various SUSY contributions among themselves 
\cite{cancel}, non universality of the soft breaking parameters at the 
unification scale \cite{non-u} and approximately degenerate heavy sfermions 
for the first two generations \cite{heavy}. 
In the presence of one of these mechanisms, large supersymmetric phases are
naturally expected and EDMs should be generally close to the experimental 
bounds.     

In this work we will study the effects of these phases in CP violation
observables as $\varepsilon_K$, $\varepsilon^\prime/\varepsilon$ and $B^0$ 
CP asymmetries. We will show that the presence of large susy phases is not 
enough to produce sizeable supersymmetric contributions to these observables.
In fact, {\it in the absence of the CKM phase, a general 
MSSM with all possible phases in the soft--breaking terms, but no new flavor 
structure beyond the usual Yukawa matrices, can never give a sizeable 
contribution to $\varepsilon_K$, $\varepsilon^\prime/\varepsilon$ or hadronic 
$B^0$ CP asymmetries}. However, as recently emphasized \cite{newflavor,non-u}, 
as soon as one introduces some new flavor structure in the soft Susy--breaking 
sector, even if the CP violating phases are flavor independent, it is indeed 
possible to get sizeable CP contribution for large Susy phases and 
$\delta_{CKM}=0$.
Then, we can rephrase our sentence above in a different way: {\it A new result 
in hadronic $B^0$ CP asymmetries in the framework of supersymmetry would be 
a direct prove of the existence of a completely new flavor structure in the 
soft--breaking terms}. This means that $B$--factories will probe the flavor 
structure of the supersymmetry soft--breaking terms even before the direct 
discovery of the supersymmetric partners \cite{flavor}. 

\section{Soft--breaking flavor structure}
\label{sec:flavor}

As announced in the introduction, the presence of new flavor structure 
in the soft--breaking terms is necessary to obtain sizeable contributions
to flavor--changing CP observables (i.e. $\varepsilon_K$, 
$\varepsilon^\prime/\varepsilon$ and hadronic $B^0$ CP asymmetries).  
To prove this we will consider any MSSM, i.e. with the minimal supersymmetric 
particle content, with general {\bf complex} soft--breaking terms, but with a 
flavor structure strictly given by the two familiar Yukawa matrices or any 
matrix strictly proportional to them. In these conditions, the most general 
structure of the soft--breaking terms at the large scale, that we call 
$M_{GUT}$, is,
\begin{eqnarray}
\label{soft}
& (m_Q^2)_{i j} = m_Q^2\ \delta_{i j}\ \ (m_U^2)_{i j} = m_U^2\ \delta_{i j} &
\nonumber \\    
&(m_D^2)_{i j} = m_D^2\ \delta_{i j}\ \ (m_L^2)_{i j} = m_L^2\ \delta_{i j}  
&\nonumber \\
& (m_E^2)_{i j} = m_E^2\ \delta_{i j}\ \ \ m_{H_1}^2 \ \ \ \ m_{H_2}^2\ \ \ \  
&\nonumber \\
& m_{\tilde{g}}\ e^{i \varphi_3} \ \ m_{\tilde{W}}\ e^{i \varphi_2} \ \ 
m_{\tilde{B}}\ e^{i \varphi_1} &\nonumber \\ 
& (A_U)_{i j}= A_U\ e^{i \varphi_{A_U}}\ (Y_U)_{i j}&
\nonumber \\ & (A_D)_{i j}= A_D\ e^{i \varphi_{A_D}}\
(Y_D)_{i j}  &\nonumber \\
& (A_E)_{i j}= A_E\ e^{i \varphi_{A_E}}\ (Y_E)_{i j}. & 
\end{eqnarray}
where all the allowed phases are explicitly written and one of them can be 
removed by an R--rotation. All other numbers or matrices in this equation 
are always real.
Notice that this structure covers, not only the CMSSM \cite{CPcons}, but also 
most of Type I string motivated models considered so far 
\cite{typeI,newcancel}, gauge mediated models \cite{gaugem}, minimal effective
supersymmetry models \cite{fully,CPbs}, etc.
 
Experiments of CP violation in the $K$ or $B$ systems only involve 
supersymmetric particles as virtual particles in the loops. This means that 
the phases in the soft--breaking terms can only appear in these experiments 
through the mass matrices of the susy particles. Then, the key point in our 
discussion will be the role played by the susy phases and the soft--breaking
terms flavor structure in the low--energy sparticle mass matrices.

It is important to notice that, even in a model with flavor--universal 
soft--breaking terms at some high energy scale, as this is the case, some 
off--diagonality 
in the squark mass matrices appears at the electroweak scale. Working on the 
basis where the squarks are rotated parallel to the quarks, the so--called 
Super CKM basis (SCKM), the squark mass matrix is not flavor diagonal at 
$M_W$. This is due to the fact that at $M_{GUT}$ there are always two 
non-trivial flavor structures, namely the two Yukawa matrices for the up and 
down quarks, not simultaneously diagonalizable. This implies that 
through RGE evolution some flavor mixing leaks into the sfermion mass matrices.
In a general Supersymmetric model, the presence of new flavor structures
in the soft breaking terms would generate large flavor mixing in the sfermion 
mass matrices. However, in the CMSSM, the two Yukawa matrices are the only 
source of flavor change. Always in the SCKM basis, any off-diagonal entry in 
the sfermion mass matrices at $M_W$ will be necessarily proportional to a 
product of Yukawa couplings.
Then, a typical estimate for the element $(i,j)$ in the $L$--$L$ down 
squark mass matrix at the electroweak scale would necessarily be (see 
\cite{CPcons} for details),
\begin{eqnarray}
\label{estimate1}
({m^2}_{LL}^{(D)})_{i j} \approx\ c\ m_Q^2\  Y^u_{i k} {Y_{j k}^u}^*,
\end{eqnarray}
with $c$ a proportionality factor between $0.1$ and 1.
This rough estimate provides the order of magnitude of the different entries
in the sfermion mass matrices. It is important to notice that if the phases 
of these elements were ${\cal O}(1)$, due to some of the phases in equation 
(\ref{soft}), we would be able to give sizeable contributions, or even 
saturate, the different CP observables \cite{MI}. Then, it is clear that the 
relevant question for CP violation experiments is the presence of imaginary 
parts in these off--diagonal entries.

As explained in \cite{CPcons,RGE}, once we have solved the Yukawa RGEs, 
the RGE equations of all soft--breaking terms are a set of linear differential 
equations. Then, they can be solved as a linear function of the initial 
conditions,
\begin{eqnarray}
\label{solution}
&m_{Q}^{2}(M_{W})=\sum_i \eta^{(\phi_i)}_{Q} m_{\phi_i}^{2} +
\sum_i \eta^{(g_i)}_{Q}\ m_{g_i}^{2}&
\nonumber \\
&+ \sum_{i\neq j} \Big(\eta^{(g_{i j})}_{Q} e^{i\varphi_{i j}} + 
\eta^{(g_{i j})\,T}_{Q} e^{- i\varphi_{i j}}\Big) m_{g_i} m_{g_j} & 
\nonumber \\
&+ \sum_{i j} \Big(\eta^{(gA_{i j})}_{Q} 
e^{i\varphi_{i A_j}} + \eta^{(gA_{i j})\,T}_{Q}  e^{- i\varphi_{i A_j}}\Big) 
m_{g_i} A_{j}& \nonumber \\ 
&+ \sum_{i\neq j} \Big(\eta^{(A_{i j})}_{Q} e^{i\varphi_{A_i A_j}}+
\eta^{(A_{i j})\,T}_{Q} e^{- i\varphi_{A_i A_j}}\Big) A_{i} A_{j}& 
\nonumber\\
&+ \sum_i \eta^{(A_i)}_{Q}  A_{i} ^{2},&
\end{eqnarray}
where $\phi_i$ refers to any scalar, $g_i$ to the different gauginos, 
$A_i$ to any tri--linear coupling and the phases $\varphi_{a b}= (\varphi_{a} 
- \varphi_{b})$. In this equation, the different $\eta$ matrices are 
$3\times3$ matrices, {\bf strictly real} and all the allowed phases have been 
explicitly written. Regarding the imaginary parts, due to the hermiticity of
the sfermion mass matrices, any imaginary part will always be associated 
to the non--symmetric part of the $\eta^{(g_i g_j)}_{Q}$, 
$\eta^{(A_i A_j )}_{Q}$ or $\eta^{(g_i A_j)}_{Q}$ matrices. 
To estimate the size of these anti--symmetric parts, we can go to the RGE 
equations for the scalar mass matrices, where we use the same conventions 
and notation as in \cite{CPcons,RGE}. 
Taking advantage of the linearity of these equations we can directly write 
the evolution of the anti--symmetric parts, $\hat{m}_{Q}^2 = 
m_{Q}^{2}-(m_{Q}^{2})^T$ as,
\begin{eqnarray}
\label{anti-sym}
&\Frac{d \hat{m}_{Q}^{2}}{d t}\ =\  - \Big( \Frac{1}{2}\ 
(\tilde{Y}_U \tilde{Y}_U^\dagger\ +\ \tilde{Y}_D 
\tilde{Y}_D^\dagger)\  \hat{m}_{Q}^{2}  \nonumber\\
&+\ \Frac{1}{2}\ \hat{m}_{Q}^{2}\ (\tilde{Y}_U \tilde{Y}_U^\dagger\ 
+\ \tilde{Y}_D \tilde{Y}_D^\dagger)  \nonumber \\
&+\ \tilde{Y}_U\ \hat{m}_{U}^{2}\ \tilde{Y}_U^\dagger\ +\ 
\tilde{Y}_D\ \hat{m}_{D}^{2}\ \tilde{Y}_D^\dagger
\nonumber\\
&+\ 2\ i\ \Im\{\tilde{A}_U \tilde{A}_U^\dagger\ +\ \tilde{A}_D 
\tilde{A}_D^\dagger\} \Big), 
\end{eqnarray}
where, due to the reality of Yukawa matrices, we have used $Y^T = Y^\dagger$, 
and following \cite{RGE} a tilde over the couplings ($\tilde{Y}$, $\tilde{A}$, 
...) denotes a re--scaling by a factor $1/(4\pi)$.
In the evolution of the $R$--$R$ squark mass matrices, $m_U^2$ and $m_D^2$, 
only one of the two Yukawa matrices, the one with equal isospin to the squarks,
is directly involved. Then, it is easy to understand that these matrices are 
in a very good approximation diagonal in the SCKM basis once you start with 
the initial conditions given in equation (\ref{soft}).
Hence, for the sake of clarity, we can safely neglect the last two terms in 
equation (\ref{anti-sym}) and 
forget about $\hat{m}_U^2$ and $\hat{m}_D^2$. However, if needed, we could 
always apply to estimate their anti--symmetric parts an analogous reasoning
as the one we show below to $\hat{m}_{Q}^{2}$.

From equation (\ref{soft}), the initial conditions for $\hat{m}_Q^2$ 
at $M_{GUT}$ are identically zero. 
This means that the only source for $\hat{m}_{Q}^{2}$ in equation 
(\ref{anti-sym}) is necessarily 
$\Im\{A_U A_U^\dagger + A_D A_D^\dagger\}$.

The next step is then to analyze the RGE for the tri--linear couplings,
\begin{eqnarray}
\label{Aurge}
&\Frac{d \tilde{A}_U}{d t}\ =\ \Frac{1}{2}\ \Big(\Frac{16}{3}\ 
\tilde{\alpha}_3 + 
3\ \tilde{\alpha}_2 + \Frac{1}{9}\ \tilde{\alpha}_1 \Big) \tilde{A}_U &
\nonumber \\
&-\  \Big(\Frac{16}{3}\ \tilde{\alpha}_3 M_3 + 3\ \tilde{\alpha}_2 M_2 +\ 
\Frac{1}{9}\ \tilde{\alpha}_1 M_1\Big) \tilde{Y}_U &
\nonumber \\
&-\  \Big( 2\ \tilde{A}_U \tilde{Y}_U^\dagger\tilde{Y}_U + 
3\ Tr(\tilde{A}_U \tilde{Y}_U^\dagger)\tilde{Y}_U & \nonumber \\
&+\ \Frac{5}{2}\ \tilde{Y}_U \tilde{Y}_U^\dagger \tilde{A}_U 
+\ \Frac{3}{2}\ Tr(\tilde{Y}_U \tilde{Y}_U^\dagger) \tilde{A}_U & \nonumber \\ 
&+\ \tilde{A}_D \tilde{Y}_D^\dagger\tilde{Y}_U +\ 
\Frac{1}{2}\ \tilde{Y}_D \tilde{Y}_D^\dagger \tilde{A}_U \Big)& 
\end{eqnarray}
with an equivalent equation for $A_D$. With the general 
initial conditions in equation (\ref{soft}), $A_U$ is complex at any scale. 
However, we are interested in the imaginary parts of $A_U A_U^\dagger$. 
At $M_{GUT}$ this combination is exactly real, but, due to different 
renormalization of different elements of the matrix, this is not
true any more at a different scale. 

However, a careful analysis of equation (\ref{Aurge}) is enough to 
convince ourselves that these imaginary parts are extremely small. 
Let us, for a moment, neglect the terms involving 
$\tilde{A}_D \tilde{Y}_D^\dagger$ or $\tilde{Y}_D \tilde{Y}_D^\dagger$ 
from the above equation. Then, the only flavor structure appearing in 
equation (\ref{Aurge}) at $M_{GUT}$ is $Y_U$. We can always go to the basis 
where $Y_U$ is diagonal and then we will have $A_U$ exactly diagonal at 
any scale. In particular this means that  $\Im\{A_U A_U^\dagger\}$ would 
always exactly vanish. A completely parallel reasoning can be applied to $A_D$ and 
$\Im\{A_D A_D^\dagger\}$. Hence, simply taking into account the flavor
structure, our conclusion is that, necessarily, any non--vanishing element of 
$\Im[A_U A_U^\dagger + A_D A_D^\dagger]$ and hence of $\hat{m}_{Q}^{2}$ 
must be proportional to $(\tilde{Y}_D \tilde{Y}_D^\dagger 
\tilde{Y}_U \tilde{Y}_U^\dagger - H.C.)$.     
So, we can expect them to be,
\begin{eqnarray}
\label{im-estimate}
& (\hat{m}_{Q}^{2})_{i<j} \approx K \left(Y_D Y_{D}^{\dagger} 
Y_U Y_{U}^{\dagger} - H.C.\right)_{i<j}& \nonumber \\
&(\hat{m}_{Q}^{2})_{1 2} \approx K \cos^{-2}\beta~ (h_{s} h_t \lambda^{5})& 
\nonumber \\
&(\hat{m}_{Q}^{2})_{1 3} \approx K \cos^{-2}\beta~ (h_{b} h_t \lambda^{3})&
\nonumber\\ 
&(\hat{m}_{Q}^{2})_{2 3} \approx K \cos^{-2}\beta~ (h_{b} h_t \lambda^{2}),&
\end{eqnarray}
where $h_{i}=m_{i}^{2}/v^2$, with $v=\sqrt{v_1^2+v_2^2}$ the vacuum 
expectation value of the Higgs, $\lambda=\sin \theta_c$ and $K$ is a 
proportionality constant that includes the effects of the running from 
$M_{GUT}$ to $M_W$. To estimate this constant we have to keep in mind that 
the imaginary parts of $A_U A_U^\dagger$ are generated through the RGE running 
and then these imaginary parts generate $\hat{m}_{Q}^{2}$ as a second order 
effect. This means that roughly $K \simeq {\cal O}(10^{-2})$ times a 
combination of initial conditions as in equation (\ref{solution}). So, 
we estimate these matrix elements to be $ (\cos^{-2} \beta \{ 10^{-12}, 
6 \times 10^{-8}, 3 \times 10^{-7}\})$ times initial conditions.
This was exactly the result we found for the $A$--$g$ terms in \cite{CPcons} 
in the framework of the CMSSM. 
In fact, now it is clear that this is the same for all the terms in 
equation (\ref{solution}), $g_i$--$A_j$,  $g_i$--$g_j$ and $A_i$--$A_j$, 
irrespectively of the presence of an arbitrary number of new phases.

As we have already said before, the situation in the $R$--$R$ matrices is 
still worse because the RGE of these matrices involves only the corresponding 
Yukawa matrix and hence, in the SCKM, they are always diagonal and real in 
extremely good approximation. 

Hence, so far, we have shown that the $L$--$L$ or $R$--$R$ squark mass 
matrices are still essentially real.
The only complex matrices, then, will still be the $L$--$R$ matrices that 
include, from the very beginning, the phases $\varphi_{A_i}$ and 
$\varphi_\mu$. Once more, the size of these entries is determined by the 
Yukawa elements with these two phases providing the complex structure.
However, this situation is not new for these more general MSSM models and it 
was already present even in the CMSSM. We can conclude, then, that the 
structure of the sfermion mass matrices at $M_W$ is not modified from the 
familiar structure already present in the CMSSM, irrespective of the presence 
of an arbitrary number of new susy phases.

In the next section we analyze the different indirect and direct CP violation 
observables in this general MSSM without new flavor structure.

\section{CP observables}

\subsection{Indirect CP violation}
\label{sec:indirect}
 
In first place, we will consider indirect CP violation both in the $K$ and 
$B$ systems. In the SM neutral meson mixing arises at one loop through the 
well--known $W$--box. However, in the MSSM, there are new contributions to 
$\Delta F=2$ processes coming from boxes mediated by supersymmetric particles. 
These are: charged Higgs boxes ($H^{\pm}$), chargino boxes ($\chi^{\pm}$) and 
gluino-neutralino boxes ($\tilde{g}$, $\chi^{0}$). ${\cal M}$--$\bar{\cal M}$ 
mixing is correctly described by the $\Delta F=2$ effective Hamiltonian, 
${\cal{H}}_{eff}^{\Delta F=2}$, which can be decomposed as,
\begin{eqnarray}
\label{DF=2}
&{\cal{H}}_{eff}^{\Delta F=2}\ =\ -\ \Frac{G_{F}^{2} M_{W}^{2}}{(2 \pi)^{2}}\ 
(V_{td}^{*} V_{tq})^{2}\ ( C_{1}(\mu)\ Q_{1}(\mu) & \nonumber \\
&\ +\ C_{2}(\mu)\  Q_{2}(\mu)\  +\ C_3(\mu)\ Q_3(\mu)).
\end{eqnarray}
With the relevant four--fermion operators given by 
\begin{eqnarray}
\label{ops}
Q_{1}&=&\bar{d}^{\alpha}_{L}\gamma^{\mu}q^{\alpha}_{L}\cdot 
\bar{d}^{\beta}_{L}\gamma_{\mu}q^{\beta}_{L},\nonumber\\ 
Q_{2}&=&\bar{d}^{\alpha}_{L}q^{\alpha}_{R}\cdot \bar{d}^{\beta}_{L}
q^{\beta}_{R},\nonumber\\ 
Q_{3}&=&\bar{d}^{\alpha}_{L}q^{\beta}_{R}\cdot \bar{d}^{\beta}_{L}
q^{\alpha}_{R},
\end{eqnarray}
where $q=s , b$ for the $K$ and $B$--systems respectively and $\alpha, 
\beta$ as color indices. In the CMSSM, these are the only three operators
present in the limit of vanishing $m_d$. The Wilson coefficients, $C_1(\mu)$,
$C_2(\mu)$ and $C_3(\mu)$, receive contributions from the different 
supersymmetric boxes,
\begin{eqnarray}
\label{wilson}
{C_{1}}(M_W)&=&{C_{1}^{W}}(M_W)+{C_{1}^{H}}(M_W)\\
&+& C_{1}^{\tilde{g},\chi^0}(M_W)+ {C_{1}^{\chi}}(M_W)\,\nonumber \\
{C_{2}}(M_W)&=&{C_{2}^{H}}(M_W) + C_{2}^{\tilde{g}}(M_W)\nonumber \\
{C_{3}}(M_W)&=& C_{3}^{\tilde{g},\chi^0}(M_W)+ {C_{3}^{\chi}}(M_W) \nonumber
\end{eqnarray} 

Both, the usual SM $W$--box and the charged Higgs box contribute to these 
operators. However, with $\delta_{CKM}=0$, these contributions do not contain 
any complex phase and hence cannot generate an imaginary part for these
Wilson coefficients.

Gluino and neutralino contributions are specifically supersymmetric. 
They involve the superpartners of quarks and gauge bosons. Here, the source of 
flavor mixing is not directly the usual CKM matrix. It is the presence of
off--diagonal elements in the sfermion mass matrices, as discussed in section
\ref{sec:flavor}. From the point of view of CP violation, we will always need
a complex Wilson coefficient. In the SCKM basis all gluino vertices are 
flavor diagonal and real. This means that in the mass insertion (MI) 
approximation, we need a complex MI in one of the sfermion lines.
Only $L$--$L$ mass insertions enter at first order in the Wilson coefficient 
${C_{1}}^{\tilde{g},\chi^0}(M_W)$. From equation (\ref{im-estimate}), the 
imaginary parts of these MI are at most ${\cal O}(10^{-6})$ for the $b$--$s$ 
transitions and smaller otherwise \cite{CPcons}. 
Comparing these values with the phenomenological bounds required to saturate 
the measured values of these processes \cite{MI} we can easily see that in
this model we are always several orders of magnitude below.

In the case of the Wilson coefficients ${C_{2}}^{\tilde{g}}(M_W)$ and 
${C_{3}}^{\tilde{g}}(M_W)$, the MI are $L$--$R$. However these MI are always 
suppressed by light masses of right handed squark, or in the case of $b$--$s$ 
transitions directly constrained by the $b \rightarrow s \gamma$ decay. Hence,
gluino boxes, in the absence of new flavor structures, can never give 
sizeable contributions to indirect CP violation processes \cite{CPcons}.

The chargino contributions to these Wilson coefficients were 
discussed in great detail in the CMSSM framework in reference \cite{CPcons}. 
In this more
general MSSM, as we have explained in section \ref{sec:flavor}, we find very 
similar results due to the absence of new flavor structure.

Basically, in the chargino boxes, flavor mixing comes explicitly from the 
CKM mixing matrix, although off--diagonality in the sfermion mass matrix 
introduces a small additional source of flavor mixing.  
\begin{eqnarray}
\label{chWC}
&C_1^\chi (M_W) = \sum_{i,j=1}^{2} \sum_{k, l=1}^{6} 
\sum_{\alpha \gamma \alpha^\prime \gamma^\prime}\ &
\nonumber \\
& \Frac{V_{\alpha^\prime d}^{*}\  
V_{\alpha q}V_{\gamma^\prime d}^{*} V_{\gamma q}}{(V_{td}^{*} V_{tq})^2}\ \ 
G^{(\alpha,k)i} G^{(\alpha^\prime,k)j*} &\nonumber \\ 
&G^{(\gamma^\prime,l)i*} G^{(\gamma,l)j}\ \ 
Y_1(z_{k}, z_{l}, s_i, s_j)& 
\end{eqnarray}
where $V_{\alpha q} G^{(\alpha,k)i}$ represent the coupling of chargino and 
squark $k$ to left--handed down quark $q$, $z_k = M_{\tilde{u}_k}^2/M_W^2$ 
and $s_i =M_{\tilde{\chi}_i}^2/M_W^2$. The explicit expressions for these 
couplings and loop functions can be found in reference \cite{CPcons}. 
$G^{(\alpha, k) i}$ are in general complex, as both $\varphi_\mu$ and 
$\varphi_{A_i}$ are present in the different mixing matrices. 
\begin{figure}
\begin{center}
\epsfxsize = 8cm
\epsffile{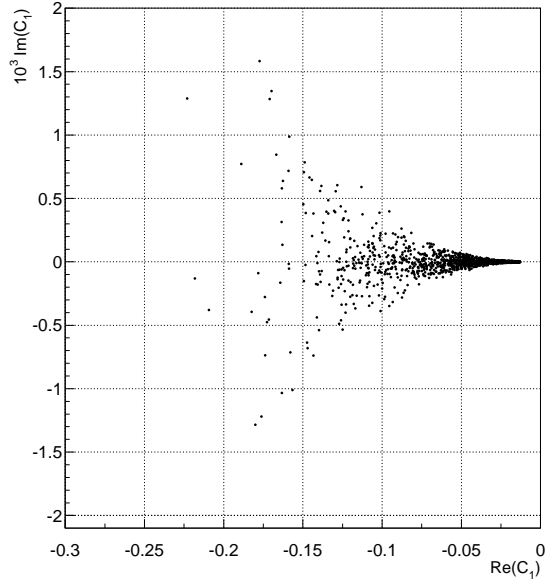}
\leavevmode
\end{center}
\caption{Imaginary and Real parts of the Wilson coefficient $C_1^\chi$ in
B mixing.}
\label{imB40}
\end{figure}

The main part of $C_1^\chi$ in equation (\ref{chWC}) will be given by pure CKM 
flavor mixing, neglecting the additional flavor mixing in the squark
mass matrix \cite{cho,branco}. This means, $\alpha= \alpha^\prime$ and
$\gamma=\gamma^\prime$. In these conditions, using the symmetry of loop 
function $Y_1(a, b, c, d)$ under the exchange of any two indices it is easy to
prove that $C_1^\chi$ would be exactly real \cite{fully}. 
This is not exactly true either in the CMSSM or in our more general MSSM, 
where there is additional flavor change in the sfermion mass matrices. 
Here, some imaginary parts appear in the $C_1^\chi$ in equation (\ref{chWC}). 
In figure \ref{imB40} we show in a scatter plot the size of imaginary 
and real parts of $C_1^\chi$ in the B system for a fixed value of 
$\tan \beta=40$. We see that this Wilson coefficient is always real up to 
a part in $10^3$. In any case, this is out of reach for the foreseen 
B--factories.

Finally, chargino boxes contribute also to the quirality changing Wilson 
coefficient $C_3^\chi(M_W)$,
\begin{eqnarray}
\label{chWCR}
&C_3^\chi (M_W) = \sum_{i,j=1}^{2} \sum_{k, l=1}^{6} \sum_{\alpha \gamma
\alpha^\prime \gamma^\prime}& \nonumber \\
&\Frac{V_{\alpha^\prime d}^{*} V_{\alpha q}V_{\gamma^\prime d}^{*} 
V_{\gamma q}}{(V_{td}^{*} V_{tq})^2}\  \Frac{m_q^2}{2 M_W^2 \cos^2 \beta}&
\nonumber \\ 
&H^{(\alpha,k)i} G^{(\alpha^\prime,k)j*} G^{(\gamma^\prime,l)i*}  
H^{(\gamma,l)j}& \nonumber \\
&Y_2(z_k, z_l, s_i, s_j)& 
\end{eqnarray} 
where $m_q/(\sqrt{2} M_W \cos \beta) \cdot V_{\alpha q} \cdot H^{(\alpha,k)i}$ 
is the coupling of chargino and squark to the right--handed down quark $q$
\cite{CPcons}.
Unlike the $C_1^\chi$ Wilson coefficient, due to the differences between $H$ 
and $G$ couplings, $C_3^\chi$ is complex even in the 
absence of intergenerational mixing in the sfermion mass matrices \cite{fully}.
Then, the presence of flavor violating entries in the up--squark mass
matrix hardly modifies the results obtained in their absence 
\cite{cho,branco,CPcons}.
In fact, in spite the presence of the Yukawa coupling squared, 
$m_q^2/(2 M^2_W \cos^2 \beta)$, this contribution could be relevant
in the large $\tan \beta$ regime. For instance, in $B^0$--$\bar{B}^0$ mixing
we have $m_b^2/(2 M^2_W \cos^2 \beta)$ that for $\tan \beta \gsim 25$ is
larger than 1 and so, it is not suppressed at all when compared with the 
$C_1^\chi$ Wilson Coefficient. This means that this contribution can be very 
important in the large $\tan \beta$ regime \cite{fully} and could have 
observable effects in CP violation experiments in the new B--factories.   
However, we will show next, that when we include the constraints coming 
from $b \rightarrow s \gamma$ these chargino contributions are also reduced 
to an unobservable level.

The chargino contributes to the $b \rightarrow s \gamma$ decay through the
Wilson coefficients ${\cal{C}}_{7}$ and ${\cal{C}}_{8}$, corresponding to the
photon and gluon dipole penguins respectively \cite{RGE,bsg,CPcons}.
In the large $\tan \beta$ regime, we can 
approximate these Wilson coefficients as \cite{CPcons},
\begin{eqnarray}
&{\cal{C}}_{7}^{\chi^{\pm}}(M_W)=\sum_{k=1}^{6}\sum_{i=1}^{2}
\sum_{\alpha, \beta =u,c,t}\ \Frac{ V_{\alpha b} V_{\beta s}^{*}}
{V_{t b} V_{t s}^{*}}& \nonumber\\
&\Frac{m_{b}}{\sqrt{2} M_W \cos \beta}\  H^{(\alpha, k) i}
{G^{*}}^{(\beta, k) i}\ \Frac{M_{\chi^{i}}}{m_{b}}\ F_{R}^{7}(z_{k},s_{i})&
\nonumber \\
&{\cal{C}}_{8}^{\chi^{\pm}}(M_W)=\sum_{k=1}^{6}\sum_{i=1}^{2}
\sum_{\alpha, \beta =u,c,t}\ \Frac{ V_{\alpha b} V_{\beta s}^{*}}
{V_{t b} V_{t s}^{*}}& \nonumber \\
&\Frac{m_{b}}{\sqrt{2} M_W \cos \beta}\  H^{(\alpha, k) i}
{G^{*}}^{(\beta, k) i}\ \Frac{M_{\chi^{i}}}{m_{b}}\ F_{R}^{8}(z_{k},s_{i})
\nonumber
\\
\label{charex}
\end{eqnarray}

Now, if we compare the chargino contributions to these Wilson coefficients
and to the coefficient $C_3$, equations (\ref{chWCR}) and (\ref{charex}), 
we can see that they are deeply related. In fact, in the approximation where
the two different loop functions involved are of the same order, we have,
\begin{eqnarray}
\label{approx}
C_3(M_W) \approx ({\cal C}_7(M_W))^2 \Frac{m_q^2}{M_W^2}
\end{eqnarray}

\begin{figure}
%\vspace{9pt}
\begin{center}
\epsfxsize = 8cm
\epsffile{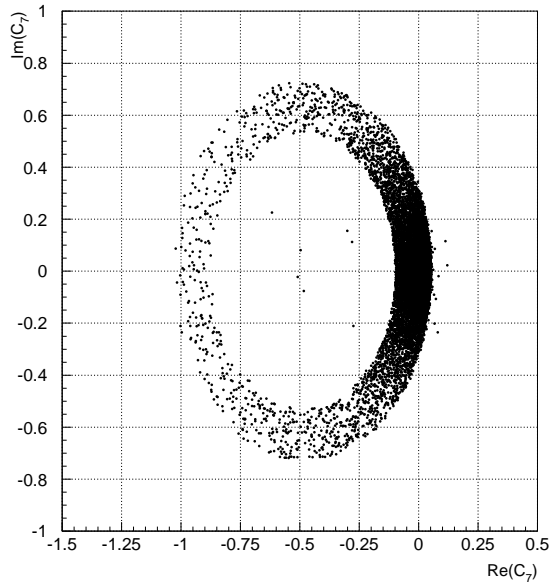}
%\framebox[55mm]{\rule[-21mm]{0mm}{43mm}}
\leavevmode
\end{center}
\caption{Experimental constraints on the Wilson Coefficient ${\cal C}_7$ }
\label{figc7}
\end{figure}

\begin{figure}
\begin{center}
\epsfxsize = 8cm
%\framebox[55mm]{\rule[-21mm]{0mm}{43mm}}
\epsffile{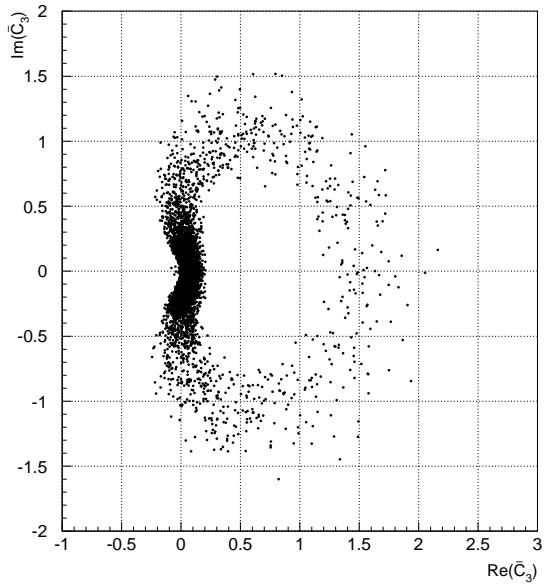}
\leavevmode
\end{center}
\caption{Allowed values for the re--scaled WC $\bar{C}_3$}
\label{figc3}
\end{figure}

In figure \ref{figc7}, we show a scatter plot of the allowed values of 
$Re({\cal C}_{7})$ versus $Im({\cal C}_{7})$ in the CMSSM for a fixed value of 
$\tan \beta = 40$ \cite{CPcons} with the constraints from the decay 
$B\rightarrow X_{s} \gamma $ taken from the reference \cite{kagan-neubert}. 
Notice that a relatively large value of $\tan \beta$, for example, 
$\tan \beta \gsim 10$, is needed to compensate the $W$ and charged Higgs 
contributions and cover the whole allowed area with positive and negative 
values. However, the shape of the plot is clearly independent of $\tan \beta$, 
only the number of allowed points and its location in the allowed area depend 
on the value considered. Then, figure \ref{figc3} shows the allowed values 
for a re--scaled Wilson coefficient $\bar{C}_3(M_W)= M^2_W/m_q^2 C_3(M_W)$ 
corresponding to the same allowed points of the susy parameter space in 
figure \ref{figc7}.
As we anticipated previously, the allowed values for $\bar{C}_3$ are close 
to the square of the values of ${\cal C}_7$ in figure \ref{figc7} slightly 
scaled by different values of the loop functions. 

We can immediately translate this result to a constraint on the size of the 
chargino contributions to $\varepsilon_{\cal M}$. 
\begin{eqnarray}
\label{epscoef}
&\varepsilon_{\cal M}\ =\ \Frac{G_F^2 M_W^2}{4 \pi^2 \sqrt{2}\ \Delta 
M_{\cal M}}\ \Frac{(V_{td} V_{tq})^2}{24}& 
\nonumber \\
&F_{\cal M}^2\  M_{\cal M}\  \Frac{M^2_{\cal M}}{m_q^2(\mu)+ 
m_d^2(\mu)}\ & \nonumber \\
&\eta_3(\mu)\  B_3(\mu)\  Im[C_3]&
\end{eqnarray}
In this expression $M_{{\cal{M}}}$, $\Delta M_{\cal M}$ and $F_{{\cal{M}}}$ 
denote the mass, mass difference and decay constant of the neutral meson 
${\cal{M}}^{0}$. The coefficient $\eta_3(\mu)= 2.93$ \cite{ciuchini} includes 
the RGE effects from $M_W$ to the meson mass scale, $\mu$, and $B_3(\mu)$ is 
the B--parameter associated with the matrix element of the $Q_3$ operator
\cite{ciuchini}.
  
For the $K$ system, using the experimentally measured value 
of $\Delta M_{K}$ we obtain,
\begin{eqnarray}
\label{epsK}
& \varepsilon_K^\chi\ =\ 1.7 \times 10^{-2}\ \Frac{m_s^2}{M_W^2}\ 
Im[\bar{C}_3] & \nonumber \\
& \approx 0.4  \times  10^{-7}\  Im[\bar{C}_3]&
\end{eqnarray}
Given the allowed values of $\bar{C}_3$ in figure \ref{figc3}, this means that
in the MSSM, even with large susy phases, chargino cannot produce a sizeable 
contribution to $\varepsilon_K$. 

The case of $B^0$--$\bar{B}^0$ mixing has a particular interest due to the 
arrival of new data from the B--factories. In fact, in the large $\tan \beta$ 
regime chargino contributions to indirect CP violation can be very important.
However, for any value of $\tan \beta$, we must satisfy the bounds from the
$b \rightarrow s \gamma$ decay. Then, if we apply these constraints to the 
$B^0$--$\bar{B}^0$ mixing,
\begin{eqnarray}
\label{epsB}
& \varepsilon_B^\chi\  =\  0.17\  \Frac{m_b^2}{M_W^2}\  Im[\bar{C}_3]&
\nonumber \\ 
& \approx 0.5 \times 10^{-3}\  Im[\bar{C}_3]&
\end{eqnarray}
where once again, with the allowed values of figure \ref{figc3} we get a very 
small contribution to CP violation in the mixing.
We must take into account that the mixing--induced CP phase, $\theta_M$, 
measurable in $B^0$ CP asymmetries, is related to $\varepsilon_B$ by 
$\theta_M=\arcsin\{2 \sqrt{2} \cdot \varepsilon_B \}$. The expected 
sensitivities on the CP phases at the B factories are around $\pm 0.1$ 
radians, so this supersymmetric chargino contribution will be absolutely 
out of reach.

\subsection{Direct CP violation}
\label{sec:direct}

To complete our analysis, we consider now direct CP violation. In this
case, the different decay processes are described by a $\Delta F=1$ 
effective Hamiltonian. A complete operator basis for these
transitions in a general MSSM involves 14 different operators \cite{MI}.
The main difference with the case of indirect CP violation is that these
operators receive contributions both from box and penguin diagrams.
Nevertheless the discussion of the presence of imaginary is completely
analogous to the case of indirect CP violation.

Once more, in the gluino case, $L$--$L$ transitions are real to a 
very good approximation, and several orders of magnitude below
the phenomenological bounds \cite{MI}. On the other hand, $L$--$R$ transitions
are suppressed by quark masses or $b \rightarrow s \gamma$ decay. This is 
always true for the squark mass matrices obtained in section \ref{sec:flavor},
and valid both for boxes and penguins.
 
Finally we are left with chargino contributions. The analysis of chargino 
boxes is exactly the same as in the previous section. In fact, even the 
Wilson coefficients are identical once we factor out the CKM elements.
Then, for the penguins, $L$--$L$ transitions are exactly real if we neglect
inter--generational mixing in the squark mass matrices. Taking into account
this small mixing we find, for the very same reasons as in the indirect CP 
violation case, that imaginary parts are far too small.
The relation of the $b \rightarrow s \gamma$ decay with the $L$--$R$ chargino 
penguins is even more transparent than before. 

However, there is still one possibility to observe the effects of the new 
supersymmetric 
phases in the absence of new flavor structure. We have seen that the reason 
for the smallness of the contributions of chargino $L$--$R$ transitions is
the experimental bound from the $B\rightarrow X_{s} \gamma$ branching ratio.
This bound makes the chirality changing transitions, although complex, too 
small to compete with $L$--$L$ transitions. Hence, in these conditions,
just the processes where only
chirality changing operators contribute (EDMs or $b\rightarrow s \gamma$), or
observables where chirality flip operators are relevant ($b\rightarrow s l^+
l^-$) can show the effects of new supersymmetric phases \cite{CPbs,CPcons}. 

\section{Conclusions}
\label{sec:conc}

To conclude we would like to summarize the possibilities of finding 
supersymmetric contributions in the different CP violation experiments.

In the presence of large supersymmetric phases,
the EDMs of the electron and the neutron must be very close to the 
experimental bounds and possibly reachable for the new generation of 
experiments. However, as we have shown in this work, the presence 
of these phases is not enough to generate a sizeable contribution to 
$\varepsilon_K$, $\varepsilon^\prime/\varepsilon$ or $B^0$ CP asymmetries. 
In this flavor changing CP observables, the presence of a completely new 
flavor structure in the soft breaking terms is a necessary ingredient
to obtain sizeable effects. Only the processes where just chirality changing 
operators contribute as $b\rightarrow s \gamma$, or processes where these 
chirality flip operators are relevant, can show the effects of the new 
supersymmetric phases.

Nevertheless, in the presence of new flavor structure in the soft 
Susy--breaking sector it is indeed possible to get sizeable CP contribution 
with large Susy phases, even with $\delta_{CKM}=0$ \cite{newflavor}. 
Then, a new result 
in hadronic $B^0$ CP asymmetries in the framework of supersymmetry would be 
a direct prove of the existence of a completely new flavor structure in the 
soft--breaking terms. And so, $B$--factories will probe the flavor 
structure of the supersymmetry soft--breaking terms even before the direct 
discovery.

\section{Acknowledgments}
We thank the organizers for the pleasant atmosphere in which this meeting
took place, D.A. Demir as the co--author of several works in which this talk is
based and S. Bertolini, T. Kobayashi and S. Khalil for enlightening 
discussions.
The work of A.M. was partially supported by the European TMR Project
``Beyond the Standard Model'' contract N. ERBFMRX CT96 0090; O.V. 
acknowledges financial support from a Marie Curie EC grant 
(TMR-ERBFMBI CT98 3087).

\end{document}